\documentclass[aps,dvipsnames]{revtex4-2}
\usepackage[english]{babel}
\usepackage{amsmath,amssymb,lineno}
\usepackage{graphicx}
\usepackage{verbatim}
\usepackage{color}
\usepackage{hyperref}
\usepackage{eqnarray,tabularx}
\usepackage{amsthm,esvect,bm,physics,cancel}
\usepackage{wrapfig}

\usepackage{pstricks,pst-circ,pstricks-add,pst-node,pst-coil,pst-eucl,pst-text,pst-fun}

\begin{document}
\title{\color{black}{Simple Pendulums in simple harmonic motion}}
\author{\color{blue}{Adel H. Alameh}}
\affiliation{\color{blue}{Lebanese University, Department of Physics, Hadath, Beirut, Lebanon}}
\date{\color{red}{\today}}
\email{adel.alameh@eastwoodcollege.com}
\begin{abstract}
 The motion of a simple pendulum in a uniform gravitational field can be described  by the solution of a second-order differential equation,
  nonlinear differential equation. In practice we solve this equation using the small angle approximation relying on students familiarity with
  simple harmonic motion.

This paper presents a straightforward method of finding the time equation of motion of a simple pendulum for small angular amplitudes, without having any recourse to solving the differential equation that governs its oscillations.\\
This method relies on finding the indefinite integral of a certain relation derived from  the conservation of mechanical energy of the system (Pendulum-Earth).
And shows no need to the mathematical complexities in which differential equations are involved.
\end{abstract}
\maketitle

\subsection*{Definition of a simple Pendulum}
A simple pendulum is an idealized body consisting of a point mass $m$ suspended by a massless, inextensible string of length $\ell$. When pulled to one side of the equilibrium position and released, the pendulum swings in a vertical plane under the influence of gravity.
\subsection*{Differential equation of the motion}
Being in the vertical position of equilibrium, the pendulum is displaced aside to the right by an angle $\theta_0$ and then released without initial velocity.
The sense of the initial displacement is considered to be positive (Fig.1). The pendulum thus oscillates under the effect of the restoring torque caused by the force of gravity $m\bm{g}$ acting on the point mass $m$. The bob ( the point mass m) is also acted upon by a force $\bm{T}$ exerted by the rod. The torque of the force $\bm{T}$ is zero since its extension meets the axis of rotation passing through the point $O$. \cite{adel1}.
\begin{center}
\includegraphics[]{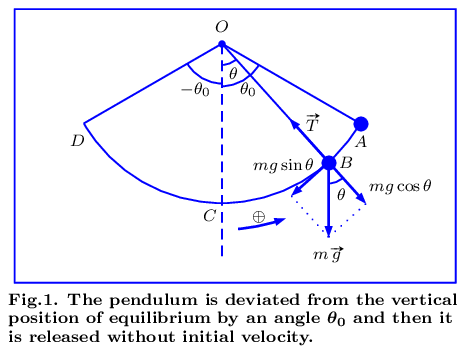}
\end{center}
\noindent The oscillations of the pendulum are governed by Newton's second law of rotational motion$\colon$
\begin{equation}\Sigma \tau_{ext}=I\theta'' \label{Newton1}\end{equation}
where $\tau$ is the torque, $I$ is the moment of inertia, and $\theta''$ is the angular acceleration of the bob. Knowing that the moment of inertia of $m$ about the axis passing through the point $O$ is $I=m\ell^2$, and that the algebraic value of the torque of $m\bm{g}$ about the point $O$ is $-mg\ell\sin\theta$\,, then equation (\ref{Newton1}) becomes
\begin{equation}m\ell^2\theta''=-mg\ell\sin\theta \label{Newton2}\end{equation}
and rearranged after canceling $m$ and $\ell$\,, it becomes
\begin{equation}\theta''+\displaystyle\frac{g}{\ell}\,\sin\theta=0 \label{Newton3}\end{equation}
 The solution of differential equation (\ref{Newton3}) can't be expressed in terms of known elementary functions. However, in the case of small angular amplitudes, $\sin\theta$ may be replaced by its first order approximation $\theta$\,. As given by taylor's expansion of $\sin\theta$ about $\theta=0$\,.
 \begin{equation}\sin\theta=\theta -\displaystyle\frac{\theta^3}{3!}+\displaystyle\frac{\theta^5}{5!}-\displaystyle\frac{\theta^7}{7!}+\cdots \label{Newton4}\end{equation}
  Thus equation (\ref{Newton3}) becomes
 \begin{equation}\theta'' +\displaystyle\frac{g}{\ell}\,\theta=0 \label{Newton5}\end{equation}
 This differential equation has the form
 \begin{equation} \theta''+\omega_0^2\, \theta=0 \label{Newton6}\end{equation}
  where \hspace{-1.25cm} \centerline{ $ \omega_0=\sqrt{\displaystyle\frac{g}{\ell}}$}
 Normally, mathematics books \cite{adel2} suggest a solution of the form $\theta=e^{bt}$ which when substituted in the differential equation (\ref{Newton6}) gives the characteristic equation $b^2+\omega_0^2=0$ and thus $b=\pm i\omega_0$, then the general solution of equation (\ref{Newton6}) is a linear combination of the two solutions. Hence
 \begin{equation}\theta=C_1e^{-i\omega_0 t} +C_2 e^{+i\omega_0 t} \label{Newton7}\end{equation}
 By making use of Euler's formula$\colon$ $e^{i\theta}=\cos\theta + i\sin\theta$, one can manipulate equation (\ref{Newton7}) to take the trigonometric form
 \begin{equation} \theta=A\cos\omega_0 t+ B \sin\omega_0 t \label{Newton8}\end{equation}
 where $A$ and $B$ are constants of respective values $A=C_1+C_2$ and $B=i(C_1-C_2)$\,.
 Furthermore equation (\ref{Newton8}) can be mathematically manipulated \cite{adel3} to take the form
 \begin{equation}\theta=C\sin(\omega_0t -\phi) \label{Newton9}\end{equation}
 where $C=\sqrt{A^2+B^2}$\,, and  $\phi=tan^{-1}\displaystyle\frac{B}{A}$\,\,$\cdot$
 \subsection*{Energy approach}
 We can apply the principle of conservation of mechanical energy on the system (Pendulum-Earth) to find the differential equation that describes the oscillations of the pendulum. (Fig.2).
 \begin{center}
\includegraphics[]{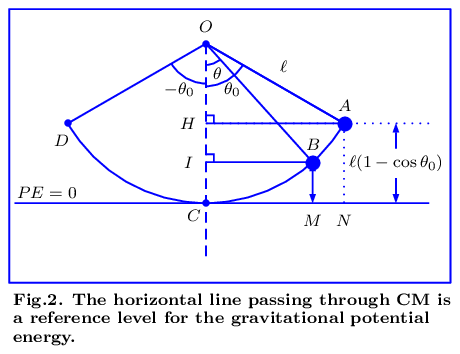}
\end{center}
 The pendulum is released without initial velocity from the angular position $\theta_0$\,. Its mechanical energy is then purely gravitational potential of value $ME=mg\ell(1-\cos\theta_0)$ where $\ell(1-\cos\theta_0)$ is the height $AN$ from the reference level. At the angular position $\theta$\,, the mechanical energy is
 $ME=\displaystyle \frac{1}{2}I\theta'^2+ mg\ell(1-\cos\theta)$\,, where $\ell(1-\cos\theta)$ is the height $BM$ from the reference level, and $I=m\ell^2$ is the moment of inertia of the point mass $m$ about the axis through the point $O$\,.
 Then, by the principle of conservation of mechanical energy in the absence of friction, one can say that
 \begin{equation} \displaystyle\frac{1}{2}m\ell^2\theta'^2 +mg\ell(1-\cos\theta)=mg\ell(1-\cos\theta_0) \label{Newton10}\end{equation}
 Now, we make the derivative of equation (\ref{Newton10}) with respect to time, and with some manipulations we get the same differential equation (\ref{Newton3})
 \begin{equation} \theta'' +\displaystyle\frac{g}{\ell}\,\sin\theta=0 \label{Newton11}\end{equation}
 and if we restrict ourselves to small angular displacements we get the same differential equation given in (\ref{Newton5})
 \begin{equation} \theta'' +\displaystyle\frac{g}{\ell}\,\theta=0 \label{Newton12}\end{equation}
 In order to find the time equation of motion, physics authors usually make use of the methods of solving differential equations  mentioned above. In addition and for the purpose of finding the period of oscillations they resort to equation (\ref{Newton10})\,. After canceling $m$ and $\ell$ and arranging, they obtain the following relation. \cite{adel4}
 \begin{equation}\theta'=\pm \sqrt{\displaystyle\frac{2g}{\ell}}\cdot\sqrt{\cos\theta -\cos\theta_0}\label{Newton13}\end{equation}
 that relation  can be written in the form
 \begin{equation}dt=\pm\sqrt{\displaystyle\frac{\ell}{2g}}\cdot \displaystyle\frac{d\theta}{\sqrt{\cos\theta-\cos\theta_0}} \label{Newton14}\end{equation}
 considering small amplitudes, $\cos\theta$ can be replaced by its first term of approximation  in its Taylor expansion.
 \begin{equation}\cos\theta =1 -\displaystyle\frac{\theta^2}{2!} + \displaystyle\frac{\theta^4}{4!}=\displaystyle\frac{\theta^6}{6!}+\cdots \label{Newton15}\end{equation}
  equation (\ref{Newton14}) thus becomes
 \begin{equation}dt=\pm\sqrt{\displaystyle\frac{\ell}{g}}\cdot \displaystyle\frac{d\theta}{\sqrt{\theta_0^2 -\theta^2}} \label{Newton16}\end{equation}
 by taking $\theta_0$ as a common factor one obtains
 \begin{equation}dt=\pm\sqrt{\displaystyle\frac{\ell}{g}} \cdot \displaystyle\frac{d\theta}{\theta_0\sqrt{1-\left(\displaystyle\frac{\theta}{\theta_0}\right)^2}} \label{Newton17}\end{equation}
 and integrating over a period, one gets
 \begin{equation}\int_0^T dt=- \sqrt{\displaystyle\frac{\ell}{g}}\int_{\theta_0}^{-\theta_0}\displaystyle\frac{d\theta}{\theta_0\sqrt{1-\left(\displaystyle\frac{\theta}{\theta_0}\right)^2}} +\sqrt{\displaystyle\frac{\ell}{g}}\int_{-\theta_0}^{\theta_0}\displaystyle\frac{d\theta}{\theta_0\sqrt{1-\left(\displaystyle\frac{\theta}{\theta_0}\right)^2}} \label{Newton18}\end{equation}
 thus
 \begin{equation}  T=\left.-\sqrt{\displaystyle\frac{\ell}{g}}\,Arcsin\left(\displaystyle\frac{\theta}{\theta_0}\right)\right|_{\theta_0}^{-\theta_0}+\left.
 \sqrt{\displaystyle\frac{\ell}{g}}\,Arcsin\left(\displaystyle\frac{\theta}{\theta_0}\right)\right|_{-\theta_0}^{\theta_0}\label{Newton19}\end{equation}
 hence
 \begin{equation} T=2\pi\sqrt{\displaystyle\frac{\ell}{g}} \label{Newton20}\end{equation}
Now, we seek to find the time equation of motion of the pendulum by a new approach that doesn't depend on solving differential equations. For that purpose we will restrict ourselves to the positive part of equation (\ref{Newton17}) and find its indefinite integral
\begin{equation}\int dt=\int \,\sqrt{\displaystyle\frac{\ell}{g}} \cdot \displaystyle\frac{d\theta}{\theta_0\sqrt{1-\left(\displaystyle\frac{\theta}{\theta_0}\right)^2}} \label{Newton21}\end{equation}
so
\begin{equation}\sqrt{\displaystyle\frac{g}{\ell}}\,t= Arcsin\left(\displaystyle\frac{\theta}{\theta_0}\right) -\phi \label{Newton22}\end{equation}
where $\phi$ is an additive constant, then by taking $\phi$ to the other side of the equation we get
\begin{equation}\sqrt{\displaystyle\frac{g}{\ell}}\,t+\phi=Arcsin\left(\displaystyle\frac{\theta}{\theta_0}\right) \label{Newton23}\end{equation}
then by operating on equation (\ref{Newton23}) by the sine function we obtain
\begin{equation}\sin\left(\sqrt{\frac{g}{\ell}}\,t +\phi\right)=\sin \left(Arcsin\left(\displaystyle\frac{\theta}{\theta_0}\right)\right)\label{Newton24}\end{equation}
and since $\left|\displaystyle\frac{\theta}{\theta_0}\right| \leq 1$\,, or in other words \, $-1\leq \displaystyle\frac{\theta}{\theta_0}\leq 1$\,, therefore the expression, $\sin\left(Arcsin\left(\displaystyle\frac{\theta}{\theta_0}\right)\right)=\displaystyle\frac{\theta}{\theta_0}$ is true.\\

\noindent Finally
\begin{equation}\theta=\theta_0 \sin\left(\sqrt{\frac{g}{\ell}}\,t +\phi\right)\label{Newton25}\end{equation}
or
\begin{equation} \theta =\theta_0\sin(\omega_0 t+\phi)\label{Newton25}\end{equation}
where \hspace{-1.25cm}\centerline{$ \omega_0=\sqrt{\displaystyle\frac{g}{\ell}}$}
If this same method is applied to the negative part of equation (\ref{Newton17}) we will get
\begin{equation}\theta =\theta_0 \cos(\omega_0 t+ \psi) \label{Newton26}\end{equation}
Both equations (\ref{Newton25}) and (\ref{Newton26}) are those of a simple harmonic motion of angular frequency $\omega_0$\,, which is equal to that of a uniform circular motion projected on a diameter, and of period
\begin{equation} T=\displaystyle\frac{2\pi}{\omega_0}=2\pi\,\sqrt{\displaystyle\frac{\ell}{g}}\label{Newton27}\end{equation}
\subsection*{Conclusion}
The indefinite integration method mentioned above applies a fortiori to all sorts of pendulums in simple harmonic motion like, elastic, torsion, and spiral pendulums, where expressions analogous to (\ref{Newton16}) emerge. Not to mention electric oscillations that take place in a series connection of a capacitor and a pure inductor where the electromagnetic energy is conserved. And in which the quantity of charges in a capacitor changes harmonically.


\begin{thebibliography}{4}
\bibitem{adel1} Adel Alameh. \emph{Phys Teach. 61, 298-301 (2023)} https://doi.org/10.1119/5.0060067.
\bibitem{adel2} L.Elsgolts. \emph{Differential Equations and the Calculus of Variations} (Mir Publishers, Moscow), p.102.
\bibitem{adel3} Murray R Speigel. \emph{Theoretical mechanics} (Schaums outline series, New York), p.92.
\bibitem{adel4} Nicolas Graber- Mutchell. \emph{arXiv:1805.00002v1}, [physics.class-ph] 28 Apr 2018.


\end{thebibliography}
 \end{document}